\documentclass[11pt]{article}

\usepackage[margin=1in]{geometry}
\usepackage[T1]{fontenc}
\usepackage[utf8]{inputenc}
\usepackage[english]{babel}
\usepackage{amsmath,amssymb,amsfonts,mathtools,amsthm}
\usepackage{graphicx}
\usepackage{hyperref}
\hypersetup{colorlinks=true,linkcolor=blue,citecolor=blue,urlcolor=blue}
\allowdisplaybreaks

\newcommand{\dd}{\mathrm{d}}
\newcommand{\Tr}{\operatorname{Tr}}

\newcommand{\Ker}{\operatorname{Ker}}
\newcommand{\Spec}{\operatorname{Spec}}
\newcommand{\Det}{\operatorname{Det}}

\newcommand{\M}{\mathcal M}
\newcommand{\D}{\mathcal D}
\newcommand{\HH}{\mathcal H}

\newcommand{\one}{\mathbf 1}
\newcommand{\ii}{\mathrm i}
\newcommand{\eps}{\epsilon}
\newcommand{\sech}{\operatorname{sech}}
\newcommand{\OmegaM}{\Omega_{\mathrm M}}
\newcommand{\partialOmegaM}{\partial\Omega_{\mathrm M}}

\newtheorem{proposition}{Proposition}

\begin{document}

\title{A Lorentzian Gribov no-pole condition for Yang--Mills theory}
\author{M. S. Guimaraes\thanks{\texttt{msguimaraes@uerj.br}}\\[4pt]
\small UERJ -- Universidade do Estado do Rio de Janeiro,\\
\small Instituto de F\'{\i}sica -- Departamento de F\'{\i}sica Te\'orica,\\
\small Rua S\~ao Francisco Xavier 524, 20550-013, Maracan\~a, Rio de Janeiro, Brazil}
\date{\today}
\maketitle

\begin{abstract}
\noindent
For over four decades, Gribov’s no-pole condition has been almost exclusively explored in Euclidean space, where the elliptic nature of the Faddeev--Popov operator provides a clear spectral boundary. In physical Minkowski spacetime, however, this operator becomes a hyperbolic wave operator, and the Euclidean positivity criterion collapses. We show that the natural Lorentzian replacement is a real-time boundary-value problem: a gauge configuration remains inside the first Gribov region as long as the Faddeev--Popov wave equation admits no source-free solutions obeying the Feynman boundary condition. For backgrounds localized in time, this condition translates to the injectivity of the negative-frequency block of the classical ghost scattering map. For stationary backgrounds, it becomes a spatial bound-state or threshold-resonance problem under Fourier transformation. Using a Wronskian identity, we prove that pure frequency mixing in stable self-adjoint time-dependent channels is structurally protected and cannot by itself produce the obstruction. While static chromomagnetic backgrounds reproduce the familiar zero-frequency horizon crossing, static chromoelectric potentials reach the horizon at finite, non-vanishing frequency—a uniquely Lorentzian phenomenon arising because $A_0$ couples directly to the ghost time derivative. We also cast the condition in Fredholm form and show that the exact restriction is a functional determinant, which the naive local continuation of the Gribov--Zwanziger action fails to reproduce, leaving the construction of a genuine local real-time action as the central open problem.
\end{abstract}

\section{Introduction}
\label{sec:introduction}

The Faddeev--Popov prescription lies at the heart of gauge theory quantization, representing the gauge-fixing Jacobian through anticommuting ghost fields \cite{Faddeev:1967fc}. Yet, this elegant construct carries a classic vulnerability: beyond perturbation theory, local gauge-fixing conditions are rarely unique. Non-Abelian gauge orbits typically pierce the gauge-fixing slice multiple times—a global topological obstruction formalized by Singer \cite{Singer:1978dk} and analyzed dynamically by Gribov in Coulomb and Landau gauges \cite{Gribov:1977wm}.
In Euclidean Landau gauge, the Faddeev--Popov operator is
\begin{equation}
\label{eq:ME}
  \M_E^{ab}[A]=-\partial_\mu D_\mu^{ab}[A],
\end{equation}
on the slice $\partial_\mu A_\mu^a=0$. Because this operator is elliptic and Hermitian on the standard functional domain, the first Gribov region can be cleanly mapped as the domain of spectral positivity:
\begin{equation}
\label{eq:EuclideanRegion}
  \Omega_E=\{A_\mu:\partial_\mu A_\mu=0,\ \M_E[A]>0\}.
\end{equation}
The boundary of this region—the Gribov horizon—is reached when the operator develops its first zero mode, signaling that the gauge-fixing slice has become parallel to a gauge orbit. This zero mode manifests physically as a pole in the ghost propagator. Gribov's no-pole condition and Zwanziger's horizon condition are two mathematically equivalent formulations of this geometric boundary, verified perturbatively and proven to all orders in Euclidean Landau gauge \cite{Gomez:2009tj,Capri:2012wx}. Crucially, Zwanziger's formulation allows the restricted path integral to be localized into a renormalizable action using auxiliary fields \cite{Zwanziger:1988jt,Zwanziger:1989mf}.

This spectral definition, however, collapses when we transition to physical Minkowski space. In Lorenz gauge,
\begin{equation}
\label{eq:LorenzGauge}
  \partial^\mu A_\mu^a=0,
\end{equation}
the Faddeev--Popov operator becomes a hyperbolic wave operator,
\begin{equation}
\label{eq:MAdefIntro}
  \M_A^{ab}=-\partial^\mu D_\mu^{ab}[A],
\end{equation}
rather than an elliptic operator. Because the spectrum of a wave operator is unbounded and typically continuous, spectral positivity loses its direct geometric meaning, and we can no longer define the Gribov region by demanding a positive-definite operator.

Gribov himself anticipated this challenge in Section 5 of his seminal work \cite{Gribov:1977wm}: in Minkowski spacetime, the spectral positivity criterion must be replaced by a boundary-value prescription. Specifically, we must impose the Feynman boundary condition on the Faddeev--Popov operator—the same prescription that dictates the real-time ghost determinant in the path integral. The problematic gauge copies are then the source-free solutions of the wave equation that behave as negative-frequency modes in the far past and as positive-frequency modes in the far future. Gribov poetically characterized this phenomenon as the classical analogue of ghost pair production in an external field.

We formalize this Lorentzian analogue by framing the no-pole condition as a boundary-value problem rather than a spectral inequality. Let $\Ker_F\M_A$ be the vector space of non-trivial, source-free solutions to the copy equation
\begin{equation}
\label{eq:copyeqintro}
  \M_A\omega=0
\end{equation}
that satisfy the Feynman boundary prescription. The first Lorentzian Gribov region is then naturally defined as:
\begin{equation}
\label{eq:OmegaIntro}
  \OmegaM^{(1)}
  =
  \{A_\mu:\partial^\mu A_\mu=0,\ \Ker_F\M_A=0\}_0 ,
\end{equation}
where the subscript $_0$ denotes the connected component containing the vacuum $A_\mu=0$.

For gauge backgrounds that are localized in time (such as pulses), the boundary-value problem \eqref{eq:copyeqintro} becomes an asymptotic ghost scattering problem. Let $\HH_\pm^{\mathrm{in}}$ and $\HH_\pm^{\mathrm{out}}$ be the positive- and negative-frequency asymptotic modes in the far past and far future, and let $S_{\mathrm{gh}}[A]$ be the classical ghost scattering matrix. The Feynman boundary condition demands that a ghost fluctuation contains only negative frequencies in the past and only positive frequencies in the future. The survival of such a mode is governed by the negative-to-negative block of the scattering matrix, $\alpha_{--}:\HH_-^{\mathrm{in}}\to\HH_-^{\mathrm{out}}$. The no-pole condition then takes the elegant algebraic form:
\begin{equation}
\label{eq:noPoleIntro}
  \Ker\alpha_{--}[A]=0.
\end{equation}
A temporally localized background thus breaches the Lorentzian horizon when $\alpha_{--}$ loses injectivity, meaning that an incoming negative-frequency ghost wave packet is completely converted or scattered away, leaving no negative-frequency component in the future. For stationary backgrounds, there is no asymptotic temporal scattering; after Fourier transforming in time, the no-pole condition \eqref{eq:OmegaIntro} manifests as a spatial bound-state or threshold-resonance condition.

We summarize the physical and formal results of this paper below:
\begin{enumerate}
\item The boundary-value kernel condition in \eqref{eq:OmegaIntro} provides a precise real-time Minkowski replacement for the traditional Euclidean spectral positivity of $\M_E$ in \eqref{eq:EuclideanRegion}.
\item For temporally localized backgrounds, this condition is equivalent to the scattering injectivity condition \eqref{eq:noPoleIntro}. For stationary backgrounds, it becomes a frequency-domain bound-state or threshold-resonance problem in a spatial channel.
\item The condition admits a Fredholm Lippmann--Schwinger form:
\begin{equation}
\label{eq:FredIntro}
  \Det_{\mathrm{Fr}}(1+G_F^0V_A)\neq0,
\end{equation}
where $G_F^0$ is the free Feynman propagator and $V_A$ is the derivative-coupled color potential. This determinant is the ghost-sector functional determinant in the background, and its zeros mark the Gribov horizon.
\item For temporally localized backgrounds, the singular values of $\alpha_{--}$ define a nonlocal flow horizon functional:
\begin{equation}
\label{eq:HflowIntro}
  H_{\mathrm{flow}}[A]
  =-\Tr_{\HH_-}\log(\alpha_{--}^\dagger[A]\alpha_{--}[A]),
\end{equation}
which acts as a dynamical barrier, diverging logarithmically as a copy mode emerges.
\item Through a Wronskian conservation identity (ghost flux conservation), we prove that pure frequency mixing in stable self-adjoint temporal channels cannot produce the obstruction at finite frequency. Simple time-dependence does not ruin gauge-fixing.
\item Static \emph{chromomagnetic} sheets reach the horizon through static, zero-frequency copy modes, reproducing the familiar Euclidean eigenvalue crossing. We demonstrate this explicitly with a smooth Pöschl--Teller magnetic sheet.
\item Static \emph{chromoelectric} wells, by contrast, cross the horizon at a \emph{finite, non-vanishing} frequency $\Omega_\star \neq 0$. Because the scalar potential $A_0$ couples directly to the time derivative in the wave operator, it acts as a frequency-dependent potential depth. The background is static, yet the copy mode it supports oscillates in time; this is a purely Lorentzian effect with no Euclidean analogue.
\item The traditional Zwanziger horizon functional admits a natural real-time Feynman-inverse continuation, but it is \emph{not} equivalent to the flow functional $H_{\mathrm{flow}}$. We show in Appendix~\ref{app:compare} that the two diverge on the horizon with different critical exponents, and that for symmetric backgrounds the naive continuation misses the horizon altogether; the exact restriction is the functional determinant, and recovering it from a local action remains open.
\end{enumerate}

Most applications of the Gribov--Zwanziger restriction originate in the Euclidean theory, relying on the analytic continuation of selected Green functions to extract real-time signatures. In contrast, our approach defines the Gribov obstruction directly as a real-time, boundary-value problem in Minkowski spacetime, prior to any continuation. To our knowledge, this is the first direct formulation of the Gribov horizon as a Minkowski Feynman-boundary problem. The closest existing real-time formulations are the canonical (Hamiltonian) analyses in Coulomb gauge \cite{Zwanziger:1998ez,Feuchter:2004mk}, which remain spatial eigenvalue problems on a fixed-time slice rather than in-out wave scattering problems in time. A related but distinct Lorentz-covariant approach to the Gribov ambiguity is Slavnov's formulation using a gauge-invariant ghost Lagrangian \cite{Slavnov:2008xz}.

We emphasize that this condition is gauge-fixed and does not, on its own, prove confinement. Its purpose is focused: it identifies the domain of classical Yang--Mills backgrounds for which the Lorenz-gauge Faddeev--Popov determinant becomes singular under the same Feynman boundary prescription that defines the real-time path integral. This provides a precise real-time counterpart of the Gribov restriction, fully aligned with the Gribov--Zwanziger physical picture in which elementary gluons exhibit a violation of reflection positivity and fail to admit a physical particle interpretation \cite{Capri:2009pv,Vandersickel:2012tz,Kondo:2019rpa}.

\section{Lorenz-gauge Faddeev--Popov operator}
\label{sec:fp}

We work in $d=4$ Minkowski space with metric
\begin{equation}
  \eta_{\mu\nu}=\operatorname{diag}(+,-,-,-),
  \qquad
  \Box=\partial^\mu\partial_\mu .
\end{equation}
The Yang--Mills gauge transformation is
\begin{equation}
\label{eq:gaugeVariation}
  \delta_\omega A_\mu^a=D_\mu^{ab}[A]\omega^b,
  \qquad
  D_\mu^{ab}[A]=\delta^{ab}\partial_\mu-g f^{abc}A_\mu^c ,
\end{equation}
with the sign convention fixed by
$[T^a,T^b]=\ii f^{abc}T^c$. The Lorenz gauge functional is
\begin{equation}
  F^a[A]=\partial^\mu A_\mu^a .
\end{equation}
Its variation is
\begin{equation}
\label{eq:FPvariation}
  \delta_\omega F^a[A]=\partial^\mu D_\mu^{ab}[A]\omega^b .
\end{equation}
We define
\begin{equation}
\label{eq:MAdef}
  \M_A^{ab}=-\partial^\mu D_\mu^{ab}[A].
\end{equation}
On the gauge slice $\partial^\mu A_\mu^a=0$ this becomes
\begin{equation}
\label{eq:MAexpanded}
  \M_A^{ab}
  =
  -\delta^{ab}\Box
  +g f^{abc}A_\mu^c\partial^\mu .
\end{equation}
Thus the ghost action is
\begin{equation}
\label{eq:ghostAction}
  S_{\mathrm{gh}}[A,\bar c,c]
  =
  \int \dd^4x\,\bar c^a\M_A^{ab}c^b .
\end{equation}
The Faddeev--Popov wave operator is not merely a tool for ghost loops; it is the mathematical sentinel guarding the uniqueness of the gauge slice. An infinitesimal gauge transformation $\omega$ deforms a Lorenz-gauge background $A_\mu$ into a redundant copy $A_\mu+D_\mu[A]\omega$ that remains on the gauge slice to first order if and only if $\omega$ lies in the kernel of this operator:
\begin{equation}
\label{eq:copyEquation}
  \M_A\omega=0 .
\end{equation}
Equation \eqref{eq:copyEquation} is therefore both the classical field equation for the ghost in the background $A_\mu$ and the infinitesimal gauge-copy equation.

To study the solutions to this wave equation, we isolate the free propagation from the background interaction by splitting the operator as:
\begin{equation}
\label{eq:Msplit}
  \M_A=\M_0+V_A,
\end{equation}
where
\begin{equation}
  \M_0^{ab}=-\delta^{ab}\Box,
  \qquad
  (V_A)^{ab}=g f^{abc}A_\mu^c\partial^\mu .
\end{equation}
Here, $\M_0$ is the free wave operator, while $V_A$ acts as a derivative-coupled color potential. Throughout this work, $G_F^0$ denotes the Green-function inverse of the free differential operator $\M_0$ with Feynman boundary conditions:
\begin{equation}
\label{eq:freeGF}
  G_F^{0,ab}(p)=
  \frac{\delta^{ab}}{p^2+\ii0}.
\end{equation}
We note that $G_F^0$ is the Green function of the wave operator itself, which differs from the standard time-ordered ghost propagator by an overall factor of $\ii$. With this convention, the Lippmann--Schwinger operator $1+G_F^0V_A$ is defined unambiguously.

\section{Boundary-value form of the Lorentzian no-pole condition}
\label{sec:bogoliubov}

Because the copy equation is a wave equation rather than an elliptic equation, the concept of a "zero mode" is mathematically incomplete until we specify boundary conditions. The natural boundary conditions are those dictated by the Feynman prescription of the ghost determinant in the real-time path integral. Let $\Ker_F\M_A$ denote the vector space of source-free solutions to
\begin{equation}
\label{eq:FkernelDef}
  \M_A\omega=0,
\end{equation}
subject to this Feynman prescription. For a background localized in time, this boundary condition requires the copy mode to be purely negative-frequency in the far past and purely positive-frequency in the far future. For a stationary background, the prescription translates to the $+\ii0$ frequency-space regularization, which manifests in the spatial channel as a bound-state or threshold-resonance condition.

\paragraph{Definition 1: Lorentzian no-pole condition.}
For any Lorenz-gauge background for which this boundary-value problem is defined, the Minkowski no-pole condition is:
\begin{equation}
\label{eq:defNoPole}
  \Ker_F\M_A=0 .
\end{equation}
The first Lorentzian Gribov region is defined as the connected component containing the vacuum:
\begin{equation}
\label{eq:OmegaM}
  \OmegaM^{(1)}
  =
  \{A_\mu:\partial^\mu A_\mu=0,\ \Ker_F\M_A=0\}_0 ,
\end{equation}
where the subscript $_0$ denotes the connected component containing $A_\mu=0$. The Lorentzian Gribov horizon is the failure boundary:
\begin{equation}
\label{eq:BoundaryM}
  \partialOmegaM
  =
  \{A_\mu:\partial^\mu A_\mu=0,\ \Ker_F\M_A\neq0\}.
\end{equation}
Crucially, this condition makes no reference to a lowest eigenvalue. Instead, it asks a purely dynamical question: does the Faddeev--Popov wave equation admit a source-free copy mode under the Feynman boundary prescription? In infinite volume, a threshold configuration can drive the horizon functionals (defined in Sec.~\ref{sec:horizon}) to divergence even in the absence of a strictly normalizable kernel vector. In finite volume, the kernel condition is literal.

The Feynman boundary condition is not an arbitrary choice. It is the prescription carried by
the ghost determinant in the in-out generating functional---the very object whose zeros we are
characterizing. Retarded or advanced prescriptions compute ghost response functions rather
than the vacuum persistence amplitude, and do not register the copy in the same way; a
genuinely in-in, Schwinger--Keldysh treatment would replace the single Feynman propagator by a
contour-ordered matrix of ghost propagators and would define its own, contour-dependent notion
of horizon. We do not pursue those variants here: the region $\OmegaM^{(1)}$ defined above is
the one relevant to the standard real-time, in-out path integral.

Assume now that $A_\mu(t,\mathbf x)$ vanishes outside a finite time interval, or decays
fast enough that free in and out decompositions exist. The free asymptotic ghost modes are
\begin{align}
\label{eq:freemodes}
  u_{\mathbf p,\lambda}^{(+)}(x)
  &=
  \frac{e^{-\ii |\mathbf p|t+\ii\mathbf p\cdot\mathbf x}}{\sqrt{2|\mathbf p|}}\,
  e_\lambda ,
\\
  u_{\mathbf p,\lambda}^{(-)}(x)
  &=
  \frac{e^{+\ii |\mathbf p|t+\ii\mathbf p\cdot\mathbf x}}{\sqrt{2|\mathbf p|}}\,
  e_\lambda ,
\end{align}
where $e_\lambda$ is a color basis vector. Let $\HH_+^{\mathrm{in}}$ and
$\HH_-^{\mathrm{in}}$ denote the positive- and negative-frequency Cauchy data at
$t\to-\infty$, and define $\HH_\pm^{\mathrm{out}}$ similarly at $t\to+\infty$.
To diagnose the propagation of these unphysical yet mathematically vital ghost fields, we equip the auxiliary asymptotic spaces with a positive-definite norm $\|\cdot\|$ defined by the sum of the modulus squared of the free mode coefficients: if $f=\sum_{\mathbf p,\lambda}c_{\mathbf p,\lambda}\,u_{\mathbf p,\lambda}^{(\pm)}$, then $\|f\|^2=\sum_{\mathbf p,\lambda}|c_{\mathbf p,\lambda}|^2$. Geometrically, this is the Klein--Gordon inner product restricted to $\HH_+$, and minus the Klein--Gordon product restricted to $\HH_-$, both of which are positive-definite on their respective subspaces. We emphasize that this norm is purely a diagnostic tool defined on the auxiliary one-particle solution space; because ghosts are strictly unphysical, this norm does not correspond to any physical Hilbert space norm.

As the ghost wave packet propagates through the time-localized gauge background $A_\mu$, the interaction maps incoming states to outgoing states. This evolution is described by the classical ghost scattering operator:
\begin{equation}
\label{eq:Sgh}
  S_{\mathrm{gh}}[A]:
  \HH_+^{\mathrm{in}}\oplus\HH_-^{\mathrm{in}}
  \longrightarrow
  \HH_+^{\mathrm{out}}\oplus\HH_-^{\mathrm{out}} .
\end{equation}
Block-decomposing this scattering operator yields:
\begin{equation}
\label{eq:Sblocks}
  S_{\mathrm{gh}}[A]=
  \begin{pmatrix}
  \alpha_{++} & \beta_{+-}\\
  \beta_{-+} & \alpha_{--}
  \end{pmatrix}.
\end{equation}
An incoming negative-frequency mode $v\in\HH_-^{\mathrm{in}}$ therefore scatters and evolves into:
\begin{equation}
\label{eq:negativeEvolution}
  v
  \longmapsto
  \beta_{+-}v+\alpha_{--}v,
  \qquad
  \beta_{+-}v\in\HH_+^{\mathrm{out}},
  \quad
  \alpha_{--}v\in\HH_-^{\mathrm{out}} .
\end{equation}
The off-diagonal block $\beta_{+-}$ measures the degree of frequency mixing. While in a physical field theory frequency mixing signals particle pair production, here it remains confined to the gauge-fixing sector: ghosts do not correspond to physical states. Instead, the mixing described by \eqref{eq:negativeEvolution} indicates that the Faddeev--Popov Jacobian is being tested along a time-dependent gauge orbit.

Gribov's no-pole condition is a much stronger requirement than the simple presence of frequency mixing ($\beta_{+-} \neq 0$). A non-vanishing $\beta_{+-}$ merely means that an incoming negative-frequency ghost disturbance is scattered partly into positive frequencies. The no-pole problem asks whether the wave equation admits a non-trivial source-free solution satisfying the Feynman endpoint conditions. This requires the future negative-frequency component to vanish completely:
\begin{equation}
\label{eq:alphaKernel}
  \exists\,v\in\HH_-^{\mathrm{in}},\quad v\neq0,
  \qquad
  \alpha_{--}[A]v=0 .
\end{equation}
Such a solution exhibits the specialized asymptotics:
\begin{equation}
\label{eq:GribovAsymptotics}
  \omega(t\to-\infty)\in\HH_-^{\mathrm{in}},
  \qquad
  \omega(t\to+\infty)\in\HH_+^{\mathrm{out}} .
\end{equation}
Thus, for temporally localized backgrounds, the no-pole condition \eqref{eq:defNoPole} simplifies to the injectivity of the negative-to-negative scattering block:
\begin{equation}
\label{eq:alphaNoPole}
  \Ker\alpha_{--}[A]=0 .
\end{equation}
The Lorentzian horizon is not triggered by the onset of frequency mixing. Rather, it is marked by the loss of injectivity of the scattering block $\alpha_{--}$, signaling the emergence of a physical Feynman-boundary copy mode.

\section{Wronskian protection}
\label{sec:wronskian}

A common concern when working in real-time is that any time-dependence in a gauge background, by mixing positive and negative frequencies, might automatically trigger a Gribov copy. Fortunately, a powerful dynamical safeguard prevents this: in stable, self-adjoint \emph{temporal} channels, the conservation of the Wronskian---physically the conservation of the auxiliary ghost flux---rules out the formation of copy modes through frequency mixing alone, at any finite frequency. The gauge slice remains structurally elastic, resisting copies under pure frequency-mixing oscillations. This is the Lorentzian analogue of the Euclidean statement that one remains strictly inside the first Gribov region until a zero mode appears.

\begin{proposition}[Wronskian protection]
\label{prop:wronskian}
Consider one charged ghost channel reduced to
\begin{equation}
\label{eq:scalarProtected}
  \ddot\phi(t)+\Omega^2(t)\phi(t)=0,
\end{equation}
with real $\Omega^2(t)$ and asymptotic limits
\begin{equation}
  \Omega(t\to-\infty)=\omega_-,
  \qquad
  \Omega(t\to+\infty)=\omega_+,
  \qquad
  \omega_\pm>0 .
\end{equation}
Let a negative-frequency in solution be expanded as
\begin{align}
  \phi(t\to-\infty)
  &\sim
  \frac{e^{+\ii\omega_- t}}{\sqrt{2\omega_-}},
\\
  \phi(t\to+\infty)
  &\sim
  \alpha\,
  \frac{e^{+\ii\omega_+ t}}{\sqrt{2\omega_+}}
  +
  \beta\,
  \frac{e^{-\ii\omega_+ t}}{\sqrt{2\omega_+}} .
\end{align}
Then
\begin{equation}
\label{eq:scalarBogIdentity}
  |\alpha|^2-|\beta|^2=1 .
\end{equation}
In particular $\alpha\neq0$.
\end{proposition}

The proof follows from the conservation of the Wronskian 
\begin{equation}
\label{eq:wronskian}
  W[\phi,\phi^*] =
  \phi\dot\phi^*-\phi^*\dot\phi .
\end{equation}
From \eqref{eq:scalarProtected} it follows that $\dot W=0$. With the flux-normalized modes above, $W=-\ii$ in the far past and
$W=-\ii(|\alpha|^2-|\beta|^2)$ in the far future. This gives
\eqref{eq:scalarBogIdentity}.

The same conclusion holds in the matrix case. If the reduced ghost equation is a stable
self-adjoint oscillator
\begin{equation}
\label{eq:matrixProtected}
  \ddot\phi+K(t)\phi=0,
  \qquad
  K(t)=K(t)^\dagger ,
\end{equation}
with positive asymptotic frequency matrices, conservation of the matrix Wronskian gives, in
flux-normalized in and out bases,
\begin{equation}
\label{eq:matrixProtectedIdentity}
  \alpha^\dagger\alpha-\beta^\dagger\beta=\one .
\end{equation}
If $\alpha v=0$ for a nonzero vector $v$, then $-\|\beta v\|^2=\|v\|^2$, which is impossible.
Thus $\Ker\alpha=0$ throughout this class. The derivation is given in
Appendix~\ref{app:wronskian}.

This is a statement about the auxiliary positive norm on asymptotic ghost mode
coefficients introduced in Sec.~\ref{sec:bogoliubov}. It is not a physical ghost Hilbert
space statement, and it does not use an indefinite ghost metric. It
tells us when $\alpha_{--}$ cannot lose injectivity.

The result fixes the role of the time-dependent examples. A Cartan pulse with real
nonzero asymptotic frequencies may have $\beta\neq0$, but that does not put the background
on the horizon. Stationary backgrounds require a different test because there is no
temporal in-out block. The chromomagnetic examples below give zero-frequency copy modes,
the direct Lorentzian image of the Euclidean lowest-eigenvalue crossing. The
chromoelectric example reaches the same kernel condition at nonzero frequency because
$A_0$ changes the spatial bound-state problem.

The assumptions of the Wronskian argument are restrictive, and it is instructive to see
exactly how they can fail. For a general $SU(2)$ Cartan background with only the color-3
components $A_\mu^3(x)$ nonzero, the charged-sector copy equation reduces to
\begin{equation}
\label{eq:CartanGeneral}
  \Box\phi_s+\ii s g\,A_\mu^3\partial^\mu\phi_s=0 ,
  \qquad
  A_\mu^3\partial^\mu=A_0^3\partial_t-\mathbf A^3\cdot\nabla ,
\end{equation}
and, for a spatially homogeneous background $A_\mu^3=A_\mu^3(t)$ with a spatial Fourier mode
$\omega=e^{\ii\mathbf p\cdot\mathbf x}\phi_s(t)\,v_s$, to the single channel
\begin{equation}
\label{eq:A0Channel}
  \ddot\phi_s+\ii s g A_0^3(t)\,\dot\phi_s
  +
  Q_s(t)\,\phi_s=0 ,
  \qquad
  Q_s(t)=\mathbf p^2+s g\,\mathbf A^3(t)\cdot\mathbf p ,
  \qquad
  s=\pm1 .
\end{equation}
Eliminating the first-derivative term brings this to the form of an effective oscillator,
\begin{equation}
\label{eq:A0EffectivePotential}
  \ddot\chi_s+
  \left[
  Q_s
  +\frac{g^2}{4}(A_0^3)^2
  -\frac{\ii}{2}s g\,\dot A_0^3
  \right]\chi_s=0 .
\end{equation}
Both reductions are carried out in Appendix~\ref{app:wronskian}.
Equation \eqref{eq:A0EffectivePotential} shows what the Wronskian argument assumes. If
$A_0^3$ is constant and $Q_s$ is real, the field redefinition is a
phase rotation and the reduced oscillator is self-adjoint. Then $\alpha_{--}$ cannot develop
a kernel. For the homogeneous channel \eqref{eq:A0Channel}, Lorenz gauge makes this
automatic: when $\mathbf A^3$ is spatially constant,
$\partial^\mu A_\mu^3=\dot A_0^3=0$, so $A_0^3$ is constant.

The imaginary term in \eqref{eq:A0EffectivePotential} can appear only after leaving this
homogeneous setting. If $\dot A_0^3\neq0$ while Lorenz gauge is maintained, the background
must also carry spatial dependence so that $\nabla\cdot\mathbf A^3$ can balance
$\dot A_0^3$. At that point the reduced temporal oscillator is no longer the whole problem.
The invariant signal is the chromoelectric field, for example
$F_{0i}^3=\partial_0A_i^3-\partial_iA_0^3$ in a Cartan sector.

A stationary chromoelectric scalar potential has no temporal scattering matrix
$\alpha_{--}$, but it still has a well-defined Feynman-boundary kernel problem after
Fourier transforming in time. The example solved in Sec.~\ref{sec:examples} reaches the
horizon through that spatial bound-state problem, at nonzero frequency. Time-dependent
chromoelectric backgrounds with both temporal and spatial structure remain to be analyzed.

\section{Fredholm form of the kernel condition}
\label{sec:feynman}

The same kernel condition can be written in Lippmann--Schwinger form. With
$\M_A=\M_0+V_A$, a Feynman-boundary solution satisfies
\begin{equation}
\label{eq:LSderivation1}
  \M_0\omega=-V_A\omega .
\end{equation}
Acting with the free Feynman inverse gives
\begin{equation}
\label{eq:LS}
  \omega=-G_F^0V_A\omega,
  \qquad
  (1+G_F^0V_A)\omega=0 .
\end{equation}
Conversely, any solution of the second equation has the Feynman boundary prescription by
construction and solves $\M_A\omega=0$. Therefore
\begin{equation}
\label{eq:kerFred}
  \Ker_F\M_A
  \simeq
  \Ker(1+G_F^0V_A).
\end{equation}

In a finite spatial volume with a ultraviolet cutoff, $1+G_F^0V_A$ reduces to a finite-dimensional matrix. The Minkowski no-pole condition then reads:
\begin{equation}
\label{eq:finiteDet}
  \det(1+G_F^0V_A)\neq0 .
\end{equation}
Here, the operator $1+G_F^0V_A$ acts on the color-valued spacetime function space $L^2(\mathbb R^{1,3})\otimes \mathfrak g$, with $G_F^0$ acting as the Feynman propagator inverse of $\M_0$.

In the thermodynamic limit of infinite volume, this finite determinant must be replaced by a Fredholm determinant. Up to a background-independent normalization, $\Det_{\mathrm{Fr}}(1+G_F^0V_A)$ is the ghost-sector functional determinant in the background $A_\mu$, that is, the ghost contribution $\Det_F\M_A/\Det\M_0$ to the gauge-fixed weight. Its vanishing marks the Gribov horizon, where the gauge-fixing description breaks down, loosely analogous to the way zeros of a partition function locate a phase boundary.

To ensure functional-analytic rigor, we assume that the gauge background is short-range in space and time, such that classical wave operators are well-defined, in accordance with standard mathematical scattering theory \cite{ReedSimon3,Yafaev1992}. We must, however, insert a word of caution regarding the definition of this determinant in $d=4$. Since the derivative-coupled potential $V_A = g f^{abc} A_\mu^c \partial^\mu$ is a first-order differential operator, the product $G_F^0 V_A$ behaves asymptotically as $A \, \partial / \partial^2 \sim A/p$ at high momentum. This mild ultraviolet falloff means that $G_F^0 V_A$ is not trace-class (it does not lie in the trace-class ideal $\mathcal S_1$), preventing us from defining the naive determinant $\Det(1+G_F^0V_A)$ directly. To make it well-defined, the determinant must be interpreted as a regularized determinant $\Det_p(1+G_F^0V_A)$ with $G_F^0V_A \in \mathcal S_p$ for a sufficiently large power $p$, or as a relative determinant defined with respect to the free evolution \cite{Simon1977}.

For the purposes of this paper, we focus on the robust and unambiguous kernel condition:
\begin{equation}
\label{eq:realCondition}
  \Ker(1+G_F^0V_A)=0 .
\end{equation}
Wherever the Fredholm determinant can be properly regularized and defined, the horizon is marked by its vanishing:
\begin{equation}
\label{eq:FredDet}
  \partialOmegaM:
  \qquad
  \Det_{\mathrm{Fr}}(1+G_F^0V_A)=0 .
\end{equation}
Within the connected component of the vacuum, this is equivalent to:
\begin{equation}
\label{eq:spectrumMinusOne}
  A\in\OmegaM^{(1)}
  \quad\Longleftrightarrow\quad
  -1\notin\Spec(G_F^0V_A).
\end{equation}

\section{Restricted path integral and horizon functionals}
\label{sec:horizon}

A formal restriction of the Lorenz-gauge real-time path integral is
\begin{equation}
\label{eq:Zrestricted}
  Z_{\OmegaM}
  =
  \int \D A\,\delta(\partial^\mu A_\mu)\,
  \Det_F \M_A\,
  \chi_{\OmegaM^{(1)}}[A]\,
  e^{\ii S_{\mathrm{YM}}[A]} .
\end{equation}
Here $\Det_F\M_A$ means the determinant with Feynman boundary conditions, and
\begin{equation}
\label{eq:chi}
  \chi_{\OmegaM^{(1)}}[A]=
  \begin{cases}
  1,& \Ker_F\M_A=0\text{ and }A\text{ belongs to the vacuum component},\\
  0,& \text{otherwise.}
  \end{cases}
\end{equation}
The characteristic function is nonlocal in time because it refers to the full
boundary-value problem. In a temporally localized background, this means the full in-out
evolution.

For temporally localized backgrounds, the singular values of $\alpha_{--}$ give the
barrier functional
\begin{equation}
\label{eq:Hflow}
  H_{\mathrm{flow}}[A]
  =
  -\Tr_{\HH_-}
  \log\!\bigl(\alpha_{--}^\dagger[A]\alpha_{--}[A]\bigr).
\end{equation}
In finite volume with a UV cutoff this is an ordinary matrix trace. In infinite volume it
should be read as a relative determinant, and we assume
\begin{equation}
\label{eq:alphaTraceClass}
  \alpha_{--}^\dagger\alpha_{--}-\one\in\mathcal S_1 .
\end{equation}
If this trace-class condition fails, one must replace $H_{\mathrm{flow}}$ by a regularized
determinant. The boundary condition $\Ker\alpha_{--}\neq0$ still has a finite-volume
meaning and gives the correct limiting criterion in the scattering problem.
In finite volume, if $s_n[A]$ are the singular values of $\alpha_{--}$, then
\begin{equation}
\label{eq:singularValues}
  H_{\mathrm{flow}}[A]
  =
  -\sum_n\log s_n^2[A].
\end{equation}
When a singular value vanishes, $H_{\mathrm{flow}}[A]\to+\infty$. Thus
\eqref{eq:Hflow} detects the time-localized horizon.

The Fredholm form gives a spacetime determinant:
\begin{equation}
\label{eq:Hdet}
  H_{\det}[A]
  =
  -\log\left|
  \Det_{\mathrm{Fr}}(1+G_F^0V_A)
  \right|^2 .
\end{equation}
Both \eqref{eq:Hflow}, when it is defined, and \eqref{eq:Hdet} diverge at a
Feynman-boundary Faddeev--Popov zero mode, but they are built on different spaces:
$H_{\mathrm{flow}}$ is a trace over the spatial asymptotic space $\HH_-$, while $H_{\det}$
is a regularized determinant on the spacetime one-particle space of
Sec.~\ref{sec:feynman}. In the scattering setting, their equality is the standard relation
between the in-out vacuum amplitude of an anticommuting field and its negative-frequency
Bogoliubov block: for the Faddeev--Popov ghost the in-out functional integral gives
\begin{equation}
  \langle\mathrm{out}|\mathrm{in}\rangle\propto\Det_F\M_A\propto\det\alpha_{--},
\end{equation}
up to a background-independent normalization \cite{FradkinGitman1991}. Under the same assumptions,
\begin{equation}
\label{eq:sameBoundary}
  H_{\mathrm{flow}}\text{ and }H_{\det}
  \text{ define the same boundary } \partial\OmegaM .
\end{equation}

A Lorentzian analogue of the Zwanziger horizon action would replace the Euclidean inverse
by the Feynman inverse \cite{Zwanziger:1988jt,Zwanziger:1989mf}. This is not part of the
definition. The boundary condition defining $\OmegaM^{(1)}$ is the Feynman-boundary kernel
condition; $\alpha_{--}$ and the Fredholm determinant are representations of it in the
settings where they exist. A local real-time Gribov--Zwanziger action would have to
reproduce this same boundary. As we show in Appendix~\ref{app:compare}, the naive Feynman
continuation of the Zwanziger functional does not: it diverges on the horizon with a
different critical exponent than $H_{\mathrm{flow}}$, and for symmetric backgrounds it can
fail to diverge at all. A faithful local action must therefore reproduce the determinant
structure of $H_{\mathrm{flow}}$, not the tree-level resolvent bilinear.

\section{Solvable examples}
\label{sec:examples}

We now illustrate these real-time boundary-value phenomena using three solvable configurations. The first is a stationary chromomagnetic background---a chromomagnetic shear layer modeled as a smooth Pöschl--Teller sheet---which reproduces the static Gribov picture: the horizon is breached by a zero-frequency, static copy mode. The second example consists of transient Cartan pulses, showing that real-time frequency mixing in a stable self-adjoint channel cannot drive the system to the horizon. The third example is a static chromoelectric well; here, the coupling of the scalar potential $A_0$ to the time derivative creates a frequency-dependent potential depth, forcing the Gribov horizon to be breached through a finite-frequency spatial bound state, with a static background but a time-oscillating copy mode.

\subsection{Static Pöschl--Teller sheet}
We model a chromomagnetic shear layer using a smooth Pöschl--Teller sheet. Take the static $SU(2)$ Cartan background:
\begin{equation}
\label{eq:staticPTBackgroundMain}
  A_0^3=0,
  \qquad
  A_y^3(x)=A\,\sech^2(\kappa x),
  \qquad
  A_\mu^{1,2}=0 .
\end{equation}
This configuration satisfies the Lorenz gauge because $A_y^3$ has no spatial $y$-dependence. In the charged color sector, we define:
\begin{equation}
  \ii\eps\,v_s=s\,v_s,
  \qquad
  s=\pm1,
  \qquad
  \eps^{12}=1 .
\end{equation}
For a spatial copy mode of the form:
\begin{equation}
  \omega^a=e^{-\ii\Omega t+\ii qy+\ii p_z z}\psi_s(x)v_s^a ,
\end{equation}
the copy equation reduces to the spatial differential equation:
\begin{equation}
\label{eq:staticPTMain}
  \psi_s''(x)+
  \left[
  \Omega^2-q^2-p_z^2-sgqA\,\sech^2(\kappa x)
  \right]\psi_s(x)=0 .
\end{equation}
This is a Schrödinger problem with a Pöschl--Teller well, whose normalizable levels are
standard \cite{PoschlTeller1933}. A bound state requires an attractive well, i.e.\
$-sgqA/\kappa^2>0$, which selects the charged channel with $sgqA<0$; the opposite channel is
repulsive and supports no copy. Writing the well strength as $\nu_s(\nu_s+1)$, the level
condition yields the dispersion relation
\begin{equation}
\label{eq:staticPTDispersionMain}
  q^2+p_z^2-\Omega_n^2
  =
  \kappa^2(\nu_s-n)^2 ,
  \qquad
  n=0,1,\dots<\nu_s ,
\end{equation}
as detailed in Appendix~\ref{app:examples}.
Writing the static problem as $\mathcal H_s(A)\psi_s=\Omega_n^2\psi_s$, the static Gribov horizon is the point where the spatial Faddeev-Popov operator develops a zero eigenvalue. Thus the critical magnetic background is obtained by setting $\Omega_n=0$. This gives the horizon condition:
\begin{equation}
\label{eq:staticPTHorizonMain}
  q^2+p_z^2=\kappa^2(\nu_s-n)^2,
  \qquad
  A_{\mathrm{hor}}^{(n)}
  =
  -\frac{s\kappa^2}{gq}\,\nu_s(\nu_s+1).
\end{equation}
For the ground state ($n=0$) with $p_z=0$ and transverse momentum $q=\kappa\nu_s>0$, the critical boundary amplitude is:
\begin{equation}
\label{eq:staticPTGroundMain}
  A_{\mathrm{hor}}
  =
  -\frac{s\kappa}{g}(\nu_s+1),
  \qquad
  \omega^a=e^{\ii\kappa\nu_s y}\sech^{\nu_s}(\kappa x)v_s^a .
\end{equation}
This represents an explicit Faddeev--Popov zero mode. It does not describe a physical ghost particle; rather, it is a stationary, redundant copy direction along the Lorenz-gauge slice. In an infinite transverse volume, the transverse plane-wave factor represents a generalized zero mode; in a finite transverse box, it becomes a strictly normalizable zero mode.

\subsection{Time-dependent Cartan pulses}
To demonstrate the robustness of the Wronskian protection, we consider transient color-field flashes modeled as homogeneous spatial pulses:
\begin{equation}
  A_i^3(t)=a_i\,\sech^2(\kappa t),
  \qquad
  A_0^3=0.
\end{equation}
For a spatial Fourier mode, the copy equation simplifies to:
\begin{equation}
  \ddot\phi_s+
  \left[
  \mathbf p^2+s g(\mathbf a\cdot\mathbf p)\sech^2(\kappa t)
  \right]\phi_s=0 .
\end{equation}
This is an exactly solvable, self-adjoint Pöschl--Teller wave equation in time. Because the temporal evolution is self-adjoint, the flux is perfectly conserved, and the Bogoliubov coefficients satisfy $|\alpha_s|^2-|\beta_s|^2=1$ for all non-vanishing momenta. Consequently, the scattering coefficient $\alpha_s$ can never vanish. The apparent singularity in the formal infrared limit $|\mathbf p|\to0$ is a threshold artifact; the coupling strength itself vanishes as $|\mathbf p|\to0$, leaving a finite, non-zero transmission limit.

A similar dynamic protection is found for a color-field step (Sauter/Rosen--Morse step) \cite{Sauter1931,RosenMorse1932}:
\begin{equation}
  A_y^3(t)=A\tanh(\kappa t),
\end{equation}
which shifts the asymptotic frequencies of the ghost modes:
\begin{equation}
  \omega_-^2=p_x^2+p_z^2+q^2-sgqA,
  \qquad
  \omega_+^2=p_x^2+p_z^2+q^2+sgqA .
\end{equation}
As long as the asymptotic frequencies remain real and positive ($\omega_\pm>0$), the flux-normalized coefficients satisfy $|\alpha_s|^2-|\beta_s|^2=1$. These solvable time-dependent examples prove that real-time frequency mixing does not equate to a Gribov copy: the gauge slice shakes but does not tear.

\subsection{Chromoelectric well}
While the magnetic configurations reproduce the zero-frequency horizon crossing of Euclidean theory, chromoelectric fields display a uniquely Lorentzian behavior: they can drive the gauge slice to the Gribov horizon at \emph{finite, non-vanishing} frequency.

Consider a static, localized Cartan scalar potential:
\begin{equation}
\label{eq:elecWellBackground}
  A_0^3(z)=-\mathcal A\,\sech^2(\kappa z),
  \qquad
  A_i^3=0 ,
\end{equation}
which satisfies Lorenz gauge trivially ($\partial^\mu A_\mu^3=\dot A_0^3=0$) and carries a localized chromoelectric field:
\begin{equation}
\label{eq:elecWellField}
  F_{0z}^3=-\partial_z A_0^3=-2\mathcal A\kappa\,\sech^2(\kappa z)\tanh(\kappa z)\neq0 ,
  \qquad F_{ij}^3=0 .
\end{equation}
Separating the copy equation using the ansatz $\omega^a=e^{-\ii\Omega t+\ii\mathbf p_\perp\cdot\mathbf x_\perp}\chi_s(z)\,v_s^a$ in the charged sector ($\ii\eps\,v_s=s\,v_s$) yields a spatial Pöschl--Teller wave equation. Because $A_0^3$ multiplies the time derivative ($\partial_t$) in the Faddeev--Popov operator, the scalar potential enters the reduced problem multiplied by the frequency $\Omega$:
\begin{equation}
\label{eq:elecWellReduced}
  \chi_s''+\Bigl[\Omega^2-\mathbf p_\perp^2+s g\,\Omega\,\mathcal A\,\sech^2(\kappa z)\Bigr]\chi_s=0 .
\end{equation}
This spatial potential has a dynamically tuned depth $sg\Omega\mathcal A$ proportional to the ghost frequency itself. At zero frequency ($\Omega=0$), the well vanishes entirely, showing that static chromoelectric fields are completely transparent to static copies. For $\Omega \neq 0$, the normalizable bound states require an attractive channel ($sg\Omega\mathcal A>0$) and satisfy:
\begin{equation}
\label{eq:elecWellHorizon}
  \kappa^2\,\nu_s(\nu_s+1)=s g\,\mathcal A\,\Omega ,
  \qquad
  \mathbf p_\perp^2-\Omega^2=\kappa^2(\nu_s-n)^2 ,
  \qquad n=0,1,\dots<\nu_s,
\end{equation}
where $\nu_s(\nu_s+1)=sg\Omega\mathcal A/\kappa^2$.

\begin{figure}[t]
\centering
\includegraphics[width=0.72\linewidth]{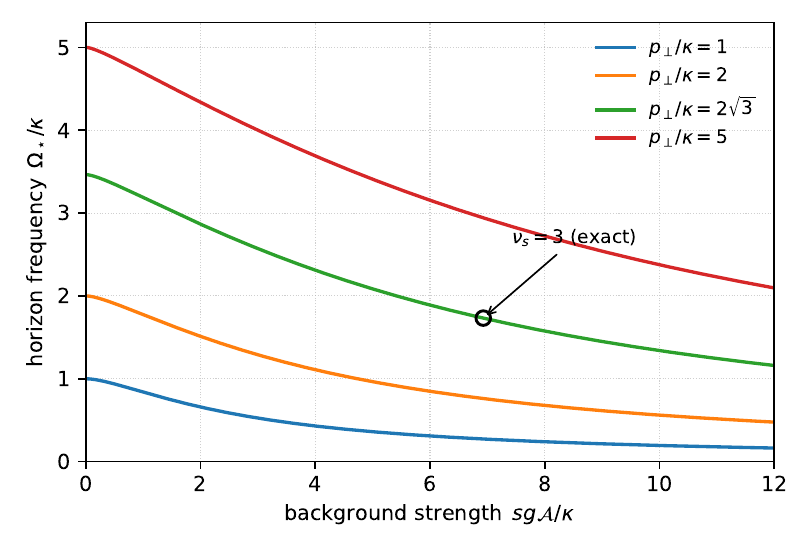}
\caption{Finite-frequency Lorentzian horizon of the static chromoelectric well. The horizon
frequency $\Omega_\star$ (ground state $n=0$), obtained by numerically solving the horizon condition \eqref{eq:elecWellExistence}, in
units of the inverse width $\kappa$, as a function of the background strength
$sg\mathcal A/\kappa$ for several transverse momenta $|\mathbf p_\perp|/\kappa$. For every
nonzero coupling the horizon lies at a strictly positive $\Omega_\star$, falling from the
threshold $\Omega_\star\to|\mathbf p_\perp|$ at weak coupling to $\Omega_\star\to0$ at strong
coupling. The open circle marks the exact reflectionless solution $\nu_s=3$ of
\eqref{eq:elecWellExact}.}
\label{fig:omegastar}
\end{figure}

For any non-zero transverse momentum, a normalizable solution exists in the attractive charged channel. For the ground state ($n=0$), writing $\beta_s=\kappa\nu_s=\sqrt{\mathbf p_\perp^2-\Omega^2}$ combines the two conditions into:
\begin{equation}
\label{eq:elecWellExistence}
  \sqrt{\mathbf p_\perp^2-\Omega^2}\,\Bigl(\sqrt{\mathbf p_\perp^2-\Omega^2}+\kappa\Bigr)
  = s g\,\mathcal A\,\Omega .
\end{equation}
As $\Omega$ ranges from $0$ to $|\mathbf p_\perp|$, the left-hand side falls from $|\mathbf p_\perp|(|\mathbf p_\perp|+\kappa)>0$ to $0$, while the right-hand side rises from $0$ to $sg\mathcal A\,|\mathbf p_\perp|>0$. By the intermediate value theorem, a unique root $\Omega_\star\in(0,|\mathbf p_\perp|)$ must exist for every transverse momentum $|\mathbf p_\perp|>0$, screening scale $\kappa>0$, and attractive potential depth $sg\mathcal A>0$. The horizon crossing does not require a discrete tuning of the background amplitude $\mathcal A$; the dependence of $\Omega_\star$ on the coupling and the transverse momentum is shown in Fig.~\ref{fig:omegastar}.

For the reflectionless value $\nu_s=3$, choosing the parameters:
\begin{equation}
\label{eq:elecWellParams}
  \kappa=\frac{|\mathbf p_\perp|}{2\sqrt3},
  \qquad
  s g\,\mathcal A=2|\mathbf p_\perp| ,
\end{equation}
we find the explicit, exact solution:
\begin{equation}
\label{eq:elecWellExact}
  \Omega_\star=\tfrac12|\mathbf p_\perp|\neq0 ,
  \qquad
  \omega^a=e^{-\frac{\ii}{2}|\mathbf p_\perp|\,t+\ii\mathbf p_\perp\cdot\mathbf x_\perp}\,
  \sech^3(\kappa z)\,v_s^a .
\end{equation}
This is an explicit, exact normalizable Faddeev--Popov copy mode at a finite, non-zero frequency. The background lies precisely on the Lorentzian Gribov horizon ($\Ker_F\M_A\neq0$), where the Feynman ghost determinant vanishes ($\Det_F\M_A=0$) and the horizon functional diverges. Just as in the magnetic sheet, the plane wave factor describes a generalized mode in infinite volume, which becomes strictly normalizable in a finite transverse box.

This chromoelectric copy differs from the magnetic examples in two crucial ways:
\begin{enumerate}
\item {\bf It has no Euclidean counterpart:} In Euclidean signature, $A_0$ and $A_i$ enter the elliptic Faddeev--Popov operator on the same footing. The frequency-dependent well depth of \eqref{eq:elecWellReduced} is therefore a purely Lorentzian phenomenon.
\item {\bf It does not violate Wronskian protection:} The protection theorem of Sec.~\ref{sec:wronskian} applies strictly to the temporal in-out scattering block. The chromoelectric well is static and has no temporal scattering; the horizon crossing occurs as a spatial bound state, where the electric potential forces the copy mode to carry a non-zero resonant frequency $\Omega_\star \neq 0$. The copy carries a spacelike invariant $\Omega_\star^2-\mathbf p_\perp^2=-\kappa^2\nu_s^2<0$, representing a localized, non-propagating copy direction rather than an asymptotic on-shell ghost particle.
\end{enumerate}

The absence of a Euclidean counterpart can be made precise by attempting the continuation
directly. Wick-rotating the frequency, $\Omega=\ii\omega_E$, turns $\Omega^2\to-\omega_E^2$ in
\eqref{eq:elecWellReduced} but renders the well depth $sg\Omega\mathcal A\to\ii\,sg\omega_E\mathcal A$
imaginary. The continued operator is no longer self-adjoint and supports no real bound state at
$\omega_E\neq0$, so the finite-frequency Lorentzian horizon is not the analytic continuation of
any Euclidean horizon. The chromomagnetic crossing, by contrast, sits at $\Omega=0=\omega_E$ and
continues trivially. The two static families therefore continue to the Euclidean problem in
opposite ways---one faithfully, the other not at all.

\section{Gauge-fixed interpretation}
\label{sec:colored}

The Faddeev--Popov determinant acts as a geometric test, verifying whether the Lorenz-gauge slice remains strictly transverse to the gauge orbits. In the real-time path integral, this transversality is evaluated under the Feynman boundary conditions. If the wave operator develops a non-trivial kernel under these conditions:
\begin{equation}
\label{eq:coloredFailure}
  \Ker_F\M_A\neq0,
\end{equation}
then the gauge-fixing map develops a null direction with the same boundary prescription as the ghost propagator. This signals that the background has crossed the boundary of the Lorentzian Gribov region \eqref{eq:OmegaM}. We emphasize that this condition is gauge-fixed and does not, on its own, serve as a gauge-invariant criterion for confinement.

For time-localized backgrounds, the no-pole condition takes on a scattering interpretation. An incoming negative-frequency ghost mode may scatter and mix with positive frequencies, but this frequency mixing is not by itself an obstruction. The horizon is reached only when the future negative-frequency component vanishes completely for a non-zero incoming mode. In a finite volume, this boundary is tracked by the smallest singular value of the negative-to-negative scattering block:
\begin{equation}
\label{eq:smin}
  s_{\min}[A]
  =
  \inf_{\|v\|=1}\|\alpha_{--}[A]v\| .
\end{equation}
The Gribov horizon is breached precisely when $s_{\min}[A]=0$.

Proposition~\ref{prop:wronskian} guarantees that this loss of injectivity cannot occur at finite frequency within any stable, self-adjoint temporal channel with positive asymptotic frequencies. Instead, the static chromoelectric well of Sec.~\ref{sec:examples} shows where the finite-frequency horizon does occur: in a spatial channel, where the scalar potential $A_0^3$ couples to the time derivative and enters the bound-state equations multiplied by the frequency. The more general time-dependent case—a chromoelectric background with simultaneous temporal and spatial dependence—remains to be explored, as its reduced evolution is no longer restricted to a simple self-adjoint oscillator.

Finally, we address a tempting physical analogy. One might draw a parallel between this horizon crossing and Schwinger pair production in an external field \cite{Schwinger:1951nm}. This comparison must, however, be kept strictly qualitative. A "ghost pair" is not a physical state; it is a classical gauge copy mode satisfying the Feynman boundary conditions. Consequently, there is no physical particle-production rate associated with this transition. Because massless ghosts lack an intrinsic mass scale, the critical threshold for any specific configuration will depend heavily on its spatial profile, color orientation, volume regulation, and infrared regularization. One should not expect a universal critical field strength or a simple exponential rate formula.

\subsection{First region versus the fundamental modular region}
It is worth being explicit about which Euclidean object $\OmegaM^{(1)}$ is the analogue of.
In Euclidean space the first Gribov region is not a fundamental domain: gauge orbits still
intersect it more than once, and the standard remedy is the fundamental modular region,
defined variationally as the set of absolute minima along each orbit of the norm
$\|A^g\|^2=\int\dd^4x\,A^g_\mu A^g_\mu$ \cite{DellAntonio:1991xt,vanBaal:1991zw}. That
construction rests on the norm being positive definite. In Minkowski signature the covariant
norm $\int\dd^4x\,A_\mu A^\mu$ is indefinite and unbounded along gauge orbits, so the
covariant minimization has no direct Lorentzian counterpart. The parallel with the rest of
this paper is close: the same indefiniteness of the metric that deprives the Euclidean
positivity condition $\M_E>0$ of meaning also deprives the modular minimization principle of
meaning. We stress that this concerns the \emph{covariant} construction only. It does not
exclude a modular region defined canonically on fixed-time slices, where the spatial norm
$\int\dd^3x\,A_iA_i$ is positive definite, nor a Coulomb-gauge or contour-dependent
alternative. Accordingly, what is constructed here is the Lorentzian analogue of the first
Gribov region, not of a fundamental modular region, and the kernel condition
$\Ker_F\M_A=0$ does not by itself guarantee the absence of finite or global copies.

\subsection{Relation to other confinement criteria}
Several other pictures of confinement --- center vortices, abelian monopoles obtained by
abelian projection, and the associated lattice criteria --- are formulated in terms of
specific classes of non-perturbative configurations rather than in terms of the gauge-fixing
determinant \cite{Greensite:2003bk,tHooft:1981ht}. It is natural to ask whether such
configurations sit on the Lorentzian horizon. For static configurations the present framework
answers this directly. If $A_0=0$ and the background is time independent, the Lorenz condition
reduces to spatial transversality, $\partial_iA_i=0$, and the Faddeev--Popov wave operator
acting on a time-independent gauge parameter $\omega(t,\mathbf x)=\psi(\mathbf x)$ collapses to
the three-dimensional operator,
\begin{equation}
\label{eq:staticBridge}
  \M_A\,\psi(\mathbf x)=\partial_iD_i[A]\,\psi(\mathbf x).
\end{equation}
Any static transverse background possessing a spatial Faddeev--Popov zero mode,
$-\partial_iD_i\psi=0$, therefore supplies a stationary $\Omega=0$ Lorentzian copy mode,
subject to the normalizability or threshold prescription discussed above. Configurations
already known to lie on a Coulomb- or Landau-gauge horizon embed into the stationary
Lorentzian horizon in this sense, and the chromomagnetic sheet of Sec.~\ref{sec:examples} is
an explicit realization: it is static, transverse, and reaches the horizon precisely at
$\Omega=0$. The corresponding statement for generic vortex or monopole configurations, which
are usually studied in other gauges and may carry time dependence or asymptotics outside the
class assumed here, is a well-posed computation of $\Ker_F\M_A$ that we have not carried out.

A separate caveat concerns the frequently used statement that configurations near the horizon
dominate. In the Euclidean theory this is a statement about a measure: the volume of the
region concentrates near its boundary in high dimension, and this is combined with the
positive weight $e^{-S}$. In real time the weight is the oscillatory phase
$e^{\ii S_{\mathrm{YM}}}$, there is no positive measure to concentrate, and it is
cancellation rather than volume that decides which configurations control the integral. The
near-horizon dominance argument therefore does not carry over as it stands, and we regard a
real-time counterpart of it as an open question rather than as a consequence of the present
construction.

\subsection{Two dimensions}
Yang--Mills theory in $1+1$ dimensions has no local propagating gauge polarizations, yet its
gauge-orbit space, holonomies, electric-flux sectors, Faddeev--Popov determinant and Gribov
regions remain nontrivial; explicit analyses exhibit Faddeev--Popov zero modes, a first Gribov
region and a fundamental modular region in that setting
\cite{Reinhardt:2008ek,Hetrick:1993ph}. There is no tension with the present framework,
because the Faddeev--Popov equation is an equation for the gauge parameter and not for
physical polarizations: the copy problem concerns gauge redundancy, and does not disappear
merely because no transverse gluon modes exist. Confinement of external color sources in two
dimensions, on the other hand, follows from Gauss's law and the resulting linear Coulomb
potential; it is kinematical and calls for no near-horizon mechanism. The coexistence of a
nontrivial Gribov problem with confinement in $1+1$ dimensions is thus perfectly consistent,
but it does not by itself establish a causal link between them. We add one technical remark
specific to our examples. The finite-frequency mechanism of Sec.~\ref{sec:examples} relies on
a transverse momentum, the localized copy mode existing because $\mathbf p_\perp^2-\Omega^2>0$
supplies spatial decay at real $\Omega$. In strict $1+1$ dimensions there is no such momentum,
so that particular real-frequency bound state is absent and the minimal setting for it is
$2+1$ dimensions. This limits the example, not the boundary-value framework.

\section{Discussion}
\label{sec:discussion}

In this work, we have proposed a direct, real-time formulation of the Gribov horizon. The first Lorentzian Gribov region is defined as the connected component containing the vacuum:
\begin{equation}
\label{eq:finalDefinition}
  \OmegaM^{(1)}
  =
  \{A:\partial^\mu A_\mu=0,\ \Ker_F(-\partial^\mu D_\mu[A])=0\}_0 ,
\end{equation}
where $\Ker_F$ denotes the kernel of the Faddeev--Popov wave operator subject to the Feynman boundary prescription. For time-localized backgrounds, this boundary-value problem translates to the injectivity of the negative-to-negative Bogoliubov scattering block:
\begin{equation}
  \Ker\alpha_{--}[A]=0 ,
\end{equation}
while in the Fredholm representation, it corresponds to the emergence of a non-trivial kernel for the Lippmann--Schwinger operator:
\begin{equation}
  \Ker(1+G_F^0V_A)=0.
\end{equation}

Our findings establish that ordinary real-time frequency mixing must not be conflated with a Gribov copy. Within a stable, self-adjoint temporal channel, the Wronskian conservation identity \eqref{eq:matrixProtectedIdentity} keeps the scattering block $\alpha_{--}$ strictly injective, even in the presence of strong frequency mixing ($\beta_{+-} \neq 0$). The Gribov horizon is reached through two distinct mechanisms in the solvable examples. Static chromomagnetic backgrounds reach the horizon at zero frequency ($\Omega=0$), reproducing the familiar spectral crossing of the Euclidean theory. In contrast, static chromoelectric wells cross the horizon at a finite, non-vanishing frequency ($\Omega_\star \neq 0$), because the scalar potential $A_0$ couples directly to the time derivative in the wave operator. This finite-frequency copy mode has no direct Euclidean counterpart, as Wick rotation places $A_0$ and $A_i$ on an identical spatial footing in the Euclidean operator.

We reiterate that this entire formulation is gauge-fixed. Ghosts remain unphysical mathematical constructs; while Becchi--Rouet--Stora--Tyutin (BRST) cohomology defines physical states perturbatively, BRST symmetry in the presence of the Gribov horizon remains a delicate open issue \cite{Dudal:2019gvn}. A Faddeev--Popov copy mode is an infinitesimal direction of gauge redundancy, and in Minkowski space, its selection is governed by the Feynman boundary prescription. Gribov's description of "ghost pair production" should be understood strictly within this geometric, gauge-fixing framework.

For backgrounds localized in time, the nonlocal barrier functional is given by the flow functional:
\begin{equation}
  H_{\mathrm{flow}}[A]
  =
  -\Tr_{\HH_-}\log(\alpha_{--}^\dagger\alpha_{--}) .
\end{equation}
By continuing the Euclidean Zwanziger horizon functional using the Feynman propagator, we can write a corresponding real-time expression:
\begin{equation}
  H_Z^M[A]
  =
  -g^2\int\dd^4x\,\dd^4y\,
  f^{adc}A_\mu^c(x)
  (\M_A^{-1})_F^{de}(x,y)
  f^{eab}A^{b\mu}(y).
\end{equation}
Although both functionals are built from the same background resolvent $\M_A^{-1}$, they are
not the same object, and the difference is instructive. In Appendix~\ref{app:compare} we show
that near the horizon $H_{\mathrm{flow}}=H_{\det}$ diverges only logarithmically, while
$H_Z^M$ develops a simple pole; and that for the symmetric backgrounds of
Sec.~\ref{sec:examples} the naive Feynman continuation $H_Z^M$ does not register the horizon
at all, because the copy mode carries quantum numbers that the background cannot source. The
exact restriction is thus a functional determinant, and a local Gribov--Zwanziger action must
reproduce that $\Tr\log$ structure rather than the tree-level bilinear $H_Z^M$.
Ultimately, translating our real-time, boundary-value horizon condition into a local, renormalizable Lorentzian Gribov--Zwanziger action remains the most compelling frontier. Resolving this challenge is not merely a mathematical exercise; it is a natural step toward the real-time spectral properties of gluons and the Minkowski-space dynamics of confinement. A useful next step will be to explore a Lorenz-gauge background that carries both a scalar potential $A_0(t, \mathbf{x})$ and explicit time dependence, allowing us to compute $\alpha_{--}$ and track its smallest singular value $s_{\min}(t)$ dynamically, and to test whether genuine time dependence sharpens the separation between $H_{\mathrm{flow}}$ and $H_Z^M$ established in Appendix~\ref{app:compare}.

\section*{Acknowledgements}

The author would like to thank the Brazilian agency Conselho Nacional de Desenvolvimento Cient\'{\i}fico e Tecnol\'ogico (CNPq) for financial support. M.~S.~Guimaraes is a CNPq researcher under contract 309793/2023-8.

\section*{Declaration on the Use of Generative AI}

The research program, derivations, and results presented here are the responsibility of the
author, who conceived the work and independently verified all of its scientific content.
Generative AI tools (OpenAI's ChatGPT/Codex and Anthropic's Claude) were used as assistive
aids: to edit and improve the language of the manuscript, to help explore and cross-check
algebra and derivations, and to assist with code generation and script automation for mathematical visualization. All visual data representations are computed directly from the analytic equations derived in the text. No
text, equation, or figure was incorporated without the author's review and verification. These
tools do not meet the criteria for authorship and are not listed as authors; the author takes
full responsibility for the entire manuscript.

\appendix

\section{Wronskian protection: derivations}
\label{app:wronskian}

This appendix collects the derivations underlying Sec.~\ref{sec:wronskian}.

\subsection{Matrix Wronskian identity}
For the stable self-adjoint oscillator \eqref{eq:matrixProtected} the relevant conserved
object is the sesquilinear form
\begin{equation}
\label{eq:matrixWronskian}
  W[\phi,\psi]=\phi^\dagger\dot\psi-\dot\phi^\dagger\psi ,
\end{equation}
defined on any two solutions. Using $\ddot\phi=-K\phi$, $\ddot\psi=-K\psi$ and
$K=K^\dagger$,
\begin{equation}
  \frac{\dd}{\dd t}W[\phi,\psi]
  =\phi^\dagger\ddot\psi-\ddot\phi^\dagger\psi
  =\phi^\dagger\bigl(K^\dagger-K\bigr)\psi=0 ,
\end{equation}
so $W$ is conserved. Take a flux-normalized basis of negative-frequency in solutions
$\phi_a$ with $\phi_a(t\to-\infty)\sim e_a\,e^{+\ii\omega_- t}/\sqrt{2\omega_-}$, written in
the eigenbasis of the (positive) asymptotic frequency matrices, and out asymptotics
\begin{equation}
  \phi_a(t\to+\infty)\sim
  \sum_b\Bigl(\alpha_{ba}\,\frac{e^{+\ii\omega_+ t}}{\sqrt{2\omega_+}}
  +\beta_{ba}\,\frac{e^{-\ii\omega_+ t}}{\sqrt{2\omega_+}}\Bigr)e_b .
\end{equation}
Evaluating the conserved matrix $\mathcal W_{ab}=W[\phi_a,\phi_b]$ in the far past gives
$\mathcal W^{-\infty}=\ii\,\one$, while in the far future the oscillating cross terms drop by
conservation and the constant part gives
$\mathcal W^{+\infty}=\ii(\alpha^\dagger\alpha-\beta^\dagger\beta)$. Equating the two yields
the identity \eqref{eq:matrixProtectedIdentity} quoted in the main text.

\subsection{Reduction of the Cartan channel}
Take an $SU(2)$ Cartan background with only the color-3 components $A_\mu^3(x)$ nonzero. In
the charged sector, with $\ii\eps\,v_s=s\,v_s$, one has
$(V_A\omega)=g\,\eps^{ab}A_\mu^3\partial^\mu\phi_s\,v_s
=-\ii s g\,A_\mu^3\partial^\mu\phi_s\,v_s$, so the copy equation $\M_A\omega=0$ for
$\omega=\phi_s v_s$ reduces to \eqref{eq:CartanGeneral}. For a spatially homogeneous
background $A_\mu^3=A_\mu^3(t)$ and a spatial Fourier mode
$\omega=e^{\ii\mathbf p\cdot\mathbf x}\phi_s(t)\,v_s$ this becomes exactly
\eqref{eq:A0Channel}, with no further terms; the coefficient of $\dot\phi_s$ is fixed by the
Faddeev--Popov operator $-\partial^\mu D_\mu$. For inhomogeneous backgrounds the spatial
gradients in \eqref{eq:CartanGeneral} add further terms.

To reach \eqref{eq:A0EffectivePotential}, write \eqref{eq:A0Channel} as
\begin{equation}
  \ddot\phi_s+P_s(t)\dot\phi_s+Q_s(t)\phi_s=0,
  \qquad
  P_s(t)=\ii s g A_0^3(t),
\end{equation}
and set
\begin{equation}
  \phi_s(t)
  =
  \exp\left[-\frac12\int^t\dd u\,P_s(u)\right]\chi_s(t),
\end{equation}
which removes the first-derivative term and leaves the effective oscillator
\eqref{eq:A0EffectivePotential}, with the standard combination
$Q_s-\tfrac14P_s^2-\tfrac12\dot P_s$.

\section{Details of the solvable reductions}
\label{app:examples}

For the static P\"oschl--Teller sheet of Sec.~\ref{sec:examples}, the reduced equation
\eqref{eq:staticPTMain} is brought to standard form by defining
\begin{equation}
\label{eq:staticPTParamsMain}
  \lambda_s=-\frac{s gqA}{\kappa^2},
  \qquad
  \rho^2=\frac{q^2+p_z^2-\Omega^2}{\kappa^2}.
\end{equation}
For an attractive well, $\lambda_s=\nu_s(\nu_s+1)>0$, the normalizable P\"oschl--Teller
levels are \cite{PoschlTeller1933}
\begin{equation}
\label{eq:staticPTLevelsMain}
  \rho_n=\nu_s-n,
  \qquad
  n=0,1,\dots<\nu_s ,
\end{equation}
which is the dispersion relation \eqref{eq:staticPTDispersionMain} quoted in the main text.
Setting $\Omega_n=0$ there gives the horizon condition
\eqref{eq:staticPTHorizonMain}, and the ground state with $p_z=0$ and $q=\kappa\nu_s$ gives
the explicit zero mode \eqref{eq:staticPTGroundMain}.

The time-dependent pulses of Sec.~\ref{sec:examples} are reduced by the same charged-sector
projection, followed by a spatial Fourier transform, which turns the copy equation into a
one-dimensional scattering problem in time; the P\"oschl--Teller and Sauter/Rosen--Morse
profiles quoted there are then the standard exactly solvable cases
\cite{PoschlTeller1933,Sauter1931,RosenMorse1932}.

\section{The flow functional versus the Feynman-continued Zwanziger functional}
\label{app:compare}

The main text introduced two functionals that both diverge on the Lorentzian horizon: the
exact flow functional $H_{\mathrm{flow}}=H_{\det}$, and the Feynman-inverse continuation
$H_Z^M$ of Zwanziger's horizon function. It is tempting to read them as two faces of one
object, much as Gribov's no-pole and Zwanziger's horizon conditions coincide in the Euclidean
theory. They are not. Both are built from the single background resolvent $\M_A^{-1}$, but
they assemble it in genuinely different ways:
$H_{\mathrm{flow}}=H_{\det}=-\log|\Det_F\M_A/\Det\M_0|^2$ is a closed ghost loop, the
logarithm of a functional determinant, whereas $H_Z^M=-g^2\!\int fA\,(\M_A^{-1})_F\,fA$ is a
single open ghost line dressed by the background---a tree-level resolvent bilinear. This
appendix makes the distinction quantitative and draws a concrete lesson from it.

\subsection{Channel reduction of $H_Z^M$}
It is convenient to first strip away the color algebra. For an $SU(2)$ Cartan background the
structure constants reduce to $f^{ad3}=\eps^{ad}$, and the charged sector is diagonalized by
the eigenvectors $\ii\eps\,v_s=s\,v_s$, so that $\eps\,v_s=-\ii s\,v_s$. The color contraction
in $H_Z^M$ then collapses to a single sum over the two charged channels, since
$\eps^{ad}(\M_A^{-1})^{de}\eps^{ea}=(-\ii s)^2\sum_s G_A^{(s)}=-\sum_s G_A^{(s)}$. This leaves
\begin{equation}
\label{eq:HZchannel}
  H_Z^M[A]=g^2\sum_{s=\pm}\int \dd^4x\,\dd^4y\;
  A_\mu^3(x)\,G_A^{(s)}(x,y)\,A^{3\mu}(y),
  \qquad
  G_A^{(s)}=\bigl(-\Box-\ii s g A_\mu^3\partial^\mu\bigr)_F^{-1},
\end{equation}
where $G_A^{(s)}$ is the Feynman resolvent of the charged-channel wave operator of
Eq.~\eqref{eq:CartanGeneral}---the very operator whose kernel defines the horizon.

\subsection{Same boundary, different critical exponents}
Approach the horizon along a one-parameter family of backgrounds, and let $\psi_0$ be the
reduced copy mode that is about to appear, with $\M_A^{(s)}\psi_0=\lambda_0\psi_0$ and a
simple crossing $\lambda_0(A)\simeq\lambda_0'\,(A_{\mathrm{hor}}-A)$, where
$\lambda_0'=\langle\psi_0|\partial_A\M_A^{(s)}|\psi_0\rangle\neq0$. The two functionals feel
this crossing very differently. The determinant carries the vanishing eigenvalue as a simple
factor, $\Det_F\M_A\propto\lambda_0$, while the resolvent carries it as a simple pole,
$\M_A^{(s)-1}=|\psi_0\rangle\langle\psi_0|/\lambda_0+(\text{regular})$. As a result,
\begin{equation}
\label{eq:logvspole}
  H_{\det}\;\sim\;-2\log|A_{\mathrm{hor}}-A|,
  \qquad
  H_Z^M\;\sim\;\frac{g^2\,|\langle A^3|\psi_0\rangle|^2}
       {\lambda_0'\,(A_{\mathrm{hor}}-A)} .
\end{equation}
Both diverge on the horizon $\partialOmegaM$, but the flow functional does so only
logarithmically, while the Zwanziger functional develops a simple pole. The two also part
company well away from the boundary: already at second order in the background,
$H_{\det}^{(2)}=\mathrm{Re}\,\Tr(G_F^0V_AG_F^0V_A)$ is a momentum loop, whereas
$H_Z^{M(2)}=g^2\sum_s\langle A^3|G_F^0|A^3\rangle$ is a tree. The only structural bridge
between them is the identity
$\partial_A H_{\det}=-2\,\mathrm{Re}\,\Tr(\M_A^{-1}\partial_A V_A)$: the gradient of the
log-determinant is a trace of the same resolvent of which $H_Z^M$ is a single matrix element.

\subsection{Overlap selection rule}
That last observation has a sharp consequence. Because $H_{\det}$ is governed by a
\emph{trace} of $\M_A^{-1}$, its divergence is automatic: the trace always contains the
critical term $\lambda_0'/\lambda_0$, which blows up whenever an eigenvalue genuinely crosses
zero. The pole of $H_Z^M$, in contrast, is weighted by how strongly the background itself
overlaps the emerging copy mode, through the factor $|\langle A^3|\psi_0\rangle|^2$; when that
overlap vanishes, $H_Z^M$ stays finite even as the determinant diverges. This is exactly what
happens for the symmetric backgrounds of Sec.~\ref{sec:examples}, where momentum conservation
forces the overlap to zero. For the Pöschl--Teller sheet the copy mode
$\psi_0\propto e^{\ii\kappa\nu_s y}\sech^{\nu_s}(\kappa x)$ carries transverse momentum
$p_y=\kappa\nu_s\neq0$, while the background $A_y^3(x)$ is independent of $y$, so the overlap
vanishes; equivalently, the coupling is proportional to $p_y$, the background decouples at the
injected momentum $p_y=0$, and \eqref{eq:HZchannel} reduces to its free, $A$-independent
value. The chromoelectric well behaves the same way: its copy mode carries
$(\Omega_\star,\mathbf p_\perp)\neq0$, while $A_0^3(z)$ is independent of $t$ and
$\mathbf x_\perp$. In both cases the flow functional diverges on the horizon, as it must,
while the naive Zwanziger functional passes through unaware.

\subsection{Lesson for a local action}
The naive Feynman continuation of the Zwanziger horizon function is therefore not a faithful
representation of the Lorentzian restriction. Where it does diverge it carries the wrong
critical exponent, and for the most symmetric---and most calculable---backgrounds it can miss
the horizon altogether. A local real-time Gribov--Zwanziger action must reproduce the
determinant, the $\Tr\log$ structure of $H_{\mathrm{flow}}$, and not merely the tree-level
bilinear $H_Z^M$. This is the precise form of the open problem raised in
Sec.~\ref{sec:discussion}.

\end{document}